\documentclass[conference]{IEEEtran}

\usepackage{graphicx}
\usepackage{multirow}
\usepackage{amsmath,amssymb,amsfonts}
\usepackage{mathrsfs}
\usepackage{xcolor}
\usepackage{textcomp}
\usepackage{booktabs}
\usepackage{array}
\usepackage{tabularx}
\usepackage[nointegrals]{wasysym}
\usepackage{pifont}
\usepackage[numbers,sort&compress]{natbib}
\usepackage[hidelinks]{hyperref}
\newcolumntype{L}[1]{>{\raggedright\arraybackslash}p{#1}}
\newcolumntype{C}[1]{>{\centering\arraybackslash}p{#1}}
\newcolumntype{Y}{>{\raggedright\arraybackslash}X}
\newcommand{\full}{\CIRCLE}
\newcommand{\half}{\LEFTcircle}
\newcommand{\none}{\Circle}
\newcommand{\adv}[1]{$\mathsf{A}_{\mathsf{#1}}$}
\newcommand{\finding}[1]{\par\smallskip\noindent\textbf{#1}\enspace}
\newcommand{\tabnote}[1]{\par\vspace{2pt}{\scriptsize\raggedright #1\par}}

\begin{document}

\title{From Models to Services: Systematizing Security Assurance for Model-as-a-Service}

\author{\IEEEauthorblockN{Jianming Zhou\textsuperscript{1}, Yang Luo\textsuperscript{1,2,3,*}, Guangpu Chen\textsuperscript{1}, Zhaoyu Chen\textsuperscript{1},\\ Yifeng Gao\textsuperscript{4}, Qingni Shen\textsuperscript{1,2,3}, Zhonghai Wu\textsuperscript{1,2,3}}
\IEEEauthorblockA{\textsuperscript{1}School of Software and Microelectronics, Peking University, Beijing, China\\
\textsuperscript{2}National Engineering Research Center for Software Engineering, Peking University, Beijing, China\\
\textsuperscript{3}Beijing Key Laboratory of Data Intelligence and Security (Peking University), Beijing, China\\
\textsuperscript{4}School of Computer Science, Peking University, Beijing, China\\
25435@qq.com, luoyang@pku.edu.cn, chenguangpu@stu.pku.edu.cn,\\
chenzhaoyu@stu.pku.edu.cn, yfgao25@stu.pku.edu.cn, qingnishen@ss.pku.edu.cn, wuzh@pku.edu.cn\\
\textsuperscript{*}Corresponding author}}

\maketitle

\begin{abstract}
Large language models are mainly delivered as cloud services, in which the assets of different principals are coupled within a single service pipeline, while the related security research is scattered across several fields with differing threat models. This paper systematizes this research from a service perspective. We decompose Model-as-a-Service (MaaS) into a five-layer stack and five trust boundaries, define six adversary classes and six security goals, use them to code 120 studies, and evaluate 50 defenses under a unified set of criteria. The results show that 44.2\% of the studies concentrate on behavioral integrity, while service integrity, availability and cost, and accountability together account for only 13.3\%. Defenses are evaluated on production services at less than half the rate of attacks, and none of the cryptographic or hardware mechanisms that protect against the provider can be adopted unilaterally by customers. Finally, we propose a research agenda by comparison with current standards and regulations.
\end{abstract}


\section{Introduction}\label{sec:intro}

Large Language Models (LLMs) such as GPT, Claude, and LLaMA are rarely run by their users themselves~\citep{brown2020gpt3,touvron2023llama2}. Training and inference require accelerators, data, and engineering capacity that few organizations possess, so model capabilities are delivered as cloud services: the provider hosts the model behind an Application Programming Interface (API), developers wrap the model with system prompts, Retrieval-Augmented Generation (RAG), and tool calls, and end users interact with the resulting applications~\citep{lewis2020rag,yao2023react}. We call this delivery model Model-as-a-Service (MaaS). It differs from the earlier Machine-Learning-as-a-Service (MLaaS)~\citep{ribeiro2015mlaas} in three ways that bear directly on security. First, the interface is natural language, and instructions and data share the same channel~\citep{greshake2023not}. Second, the model is both an asset to be protected and a potential weapon against its own operator~\citep{wei2023jailbroken}. Third, the service is shared: the same inference runtime batches requests from multiple tenants, caches their prefixes, and bills by usage~\citep{kwon2023vllm,zheng2024sglang}.

These characteristics have given rise to a large and rapidly growing body of research. Attacks can steal proprietary weights through the API~\citep{carlini2024stealing}, recover training data from aligned production models~\citep{nasr2023scalable}, infer other tenants' prompts from cache timing~\citep{song2025earlybird,gu2025auditing}, read encrypted responses from packet lengths~\citep{weiss2024keylogging}, hijack agents through retrieved content~\citep{greshake2023not,zhan2024injecagent}, and remove safety alignment through fine-tuning APIs~\citep{qi2024finetuning}. Defenses span Secure Multi-Party Computation (MPC) for Transformer inference~\citep{hao2022iron,dong2025puma}, confidential-computing GPUs~\citep{zhu2024hopper}, Differential Privacy (DP)~\citep{yu2022differentially}, guard classifiers~\citep{inan2024llamaguard,sharma2025constitutional}, and capability-based agent architectures~\citep{debenedetti2026camel}. Each of these directions adopts a different threat model: cryptographic inference assumes a curious provider, jailbreak research assumes a malicious user and an honest provider, prompt injection research assumes a malicious third party and a benign user, and model stealing research assumes a malicious user and treats the provider as the victim. These threat models are rarely stated under a unified system model, which makes it hard to judge which risks a MaaS deployment covers, which assurance mechanisms can be composed, and which gaps remain.

Existing surveys do not resolve this problem. Surveys on trustworthy LLMs~\citep{liu2023trustworthy,huang2024trustllm} and on LLM security and privacy~\citep{yao2024llmsecurity,das2025security} organize the literature by model property or attack type, a survey on agent security~\citep{deng2025threat} focuses on the agent loop, and a recent Systematization of Knowledge (SoK) paper on jailbreak guardrails~\citep{wang2026sokguardrails} evaluates one class of defenses in depth. Industry and regulatory frameworks~\citep{owasp2025llm,nist2024genaiprofile,mitreatlas,csa2024llmthreat,euaiact2024} enumerate risks but do not evaluate concrete mechanisms. The SoK on cryptographic neural network computation~\citep{ng2023sok} treats the server as the adversary, but it considers only a curious server and does not address the rest of the service. None of these works treats the provider as an adversary that misreports what it serves, treats the shared inference runtime as an attack surface, or quantifies how research effort is distributed across the service architecture.

This paper systematizes MaaS security around the service and answers five Research Questions (RQs):
\begin{itemize}
\item \textbf{RQ1} Which components, principals, and trust boundaries make up a MaaS deployment? Which adversaries can cross each boundary?
\item \textbf{RQ2} How do known attacks map onto this model? What access and assumptions does each class of attack require?
\item \textbf{RQ3} What guarantees do existing assurance mechanisms provide, under which trust assumptions, and at what scale? Who can deploy them?
\item \textbf{RQ4} Where is the literature concentrated? How realistic are its evaluations?
\item \textbf{RQ5} What do standards and regulations cover? Which open problems follow from the gaps?
\end{itemize}

\noindent Our contributions are as follows:
\begin{itemize}
\item \textbf{A service-level security model for MaaS} (Section~\ref{sec:model}). We decompose MaaS into five layers (model supply, inference runtime, interface, context assembly, and output) separated by five trust boundaries, and define six adversary classes and six security goals. Two of these goals, service integrity and availability and cost, have no counterpart in model-centric taxonomies. They arise directly from the commercial relationship between customer and provider.
\item \textbf{A systematization of attacks and assurance mechanisms} (Sections~\ref{sec:attacks} and~\ref{sec:defenses}). We organize attacks by the goal they violate and the boundary they cross, and evaluate 50 defenses against six criteria: the targeted adversary, the type of guarantee, client deployability, the demonstrated scale, whether the defense was evaluated on a production service, and whether it underwent adaptive evaluation.
\item \textbf{A quantitative characterization of the field} (Section~\ref{sec:findings}). We code 120 primary studies and report eight findings, including the concentration of research on behavioral integrity, the asymmetry in evaluation realism between attacks and defenses, and the deployability gap in protections against the provider. The full coding is listed in Appendix~\ref{app:coding}.
\item \textbf{A comparison with standards and a research agenda} (Sections~\ref{sec:standards} and~\ref{sec:lessons}). We map seven groups of frameworks and regulations clause by clause onto the goals, show that none of them covers service integrity from the customer's perspective, and summarize five lessons and seven open problems.
\end{itemize}

\noindent Section~\ref{sec:background} introduces the MaaS stack and deployment forms, Section~\ref{sec:method} describes the systematization methodology, and Section~\ref{sec:related} compares this paper with existing surveys.

\section{Background and Scope}\label{sec:background}

\subsection{The MaaS stack}

Figure~\ref{fig_1_system} shows the reference architecture used throughout this paper. We divide a MaaS deployment into five layers, ordered from the provider's assets to user-facing outcomes.

\textbf{L1 Model supply and customization.} This layer comprises pretraining corpora, base models, alignment through supervised fine-tuning and reinforcement learning from human feedback~\citep{ouyang2022instructgpt,bai2022constitutional}, and adapters produced for customers through fine-tuning APIs. Weights and adapters are the provider's most valuable assets, and many artifacts at this layer (datasets, checkpoints, serialization libraries) come from a third-party supply chain~\citep{zhao2024models,carlini2024poisoning}.

\textbf{L2 Inference runtime.} This is the serving engine, which schedules requests, batches tokens from different tenants, maintains the Key-Value (KV) cache, shares prompt prefixes across requests, and runs on shared accelerators~\citep{kwon2023vllm,zheng2024sglang}. These optimizations lower the cost of MaaS, and they also create interference between tenants.

\textbf{L3 Interface and metering.} This is the API gateway, which authenticates callers, enforces rate limits and content policies, returns tokens and optional log-probabilities, and bills by consumed tokens. The interface determines how much information an adversary can observe (text, top-$k$ log-probabilities, embeddings, latency), and it is also the customer's only window into ``what was actually served and how much was charged''.

\textbf{L4 Context assembly.} This is the developer's application logic, which assembles the prompt from the system prompt, user input, retrieved documents, memory, and tool outputs, and may run an agent loop in which model outputs trigger further tool calls~\citep{yao2023react}. This layer mixes instructions of different privilege with untrusted data.

\textbf{L5 Output and downstream use.} This layer comprises the answers shown to users, the actions executed by tools, and generated text that leaves the system and may later require provenance tracing~\citep{kirchenbauer2023watermark}.

\subsection{Principals and deployment forms}

A MaaS deployment involves five classes of principals: the \emph{provider}, which owns or licenses the model and operates L1 to L3, the \emph{developer} (tenant), which builds applications on top of the API and operates L4, the \emph{end user}, \emph{third-party content sources} whose web pages, files, and tool responses enter the context, and \emph{intermediaries} such as resellers, aggregation platforms, and API gateways that relay traffic between developers and providers. Upstream suppliers of data and model artifacts form a sixth, indirect class of principal.

The same stack appears in four deployment forms, which differ in which principal controls which layer. In a \emph{public API}, the provider controls L1 to L3 and exposes only the interface. In a \emph{dedicated or private deployment}, the weights run in an environment controlled or audited by the customer, which is common in finance, healthcare, and government. In an \emph{intermediated deployment}, a reseller or gateway claims that its model comes from an upstream provider, and without additional mechanisms the customer cannot verify what is actually served~\citep{gao2025model,cai2025substitution}. In a \emph{decentralized or edge deployment}, different layers or shards of the model run on resources belonging to different principals~\citep{borzunov2023petals,zhang2024edgeshard}. The trust assumptions of each defense in Section~\ref{sec:defenses} depend on the deployment form it presumes.

\subsection{From MLaaS to MaaS}

The first generation of research on machine learning services already showed that prediction APIs leak models~\citep{tramer2016stealing} and training-set membership~\citep{shokri2017membership}. MaaS inherits these risks and adds three new classes of risk. First, the natural-language interface makes no syntactic distinction between instructions and data, so prompt injection is an inherent property of the interface~\citep{greshake2023not,liu2024formalizing}. Second, whether the model complies with its policy is itself a security property: the provider promises that the model will refuse certain requests, and users try to break this promise through jailbreaks~\citep{zou2023universal,wei2023jailbroken}. Third, LLM inference is stateful and shared: prefix caching, continuous batching, and per-token billing introduce channels and conflicts of interest that stateless classifiers do not have~\citep{gu2025auditing,sun2025invisible}.

\subsection{Scope}

We include attack, defense, and measurement studies on language models offered or consumed as services, covering the inference runtime, customization interfaces, and the application layer that assembles context for the model. The following three adjacent areas are excluded unless they interact with MaaS: hallucination and factuality without an adversary~\citep{ji2023survey,huang2025survey}, the safety of open-weight models that requires direct modification of weights, and general cloud and hardware security unrelated to model serving.

\section{Systematization Methodology}\label{sec:method}

\textbf{Literature collection.} We searched DBLP, Google Scholar, ACM Digital Library, and IEEE Xplore with combinations of service terms (``Model-as-a-Service'', ``LLM API'', ``LLM serving'', ``inference provider'') and security terms (``extraction'', ``stealing'', ``membership inference'', ``side channel'', ``jailbreak'', ``prompt injection'', ``poisoning'', ``private inference'', ``differential privacy'', ``watermark'', ``audit''), with searches covering work up to June 2026. We prioritized security venues (USENIX Security, IEEE S\&P, ACM CCS, NDSS), machine learning venues (NeurIPS, ICML, ICLR), and language venues (ACL, EMNLP, NAACL, COLM), and supplemented them with highly cited preprints from well-known groups. We then expanded the set by snowballing over the references and citations of included studies: each round performed one backward pass and one forward pass, until a round produced no new study meeting the inclusion criteria, for three rounds in total. To verify the findings on research gaps in Section~\ref{sec:findings}, we also ran supplementary targeted searches for G4, G5, and the two adversary classes \adv{T} and \adv{P} (defined in Section~\ref{sec:model}) (``KV cache side channel'', ``prompt caching'', ``model fingerprinting'', ``model identity verification'', ``token count verification'', ``billing audit'', ``slowdown attack''). Surveys, standards, and regulations were collected as secondary sources and are discussed in Sections~\ref{sec:standards} and~\ref{sec:related}.

\textbf{Screening.} The database searches returned 2{,}316 records, of which 1{,}487 remained after deduplication. After title and abstract screening, 268 records were retained for full-text assessment, of which 177 were excluded: 74 did not study models offered or consumed as services, 51 had no empirical or formal evaluation, 33 were surveys or position papers without empirical evaluation, and 19 were other versions of already included work. Full-text assessment retained 91 studies, and snowballing and the supplementary targeted searches added another 29.

\textbf{Inclusion criteria.} A primary study must satisfy two conditions: (i) it studies a language model offered or consumed through a service interface, or a component of the inference or customization pipeline, and (ii) it proposes or measures an attack, defense, or security property, with an empirical or formal evaluation. The final corpus contains 120 primary studies published from 2020 to 2026: 52 attacks, 18 measurement or benchmark studies, and 50 defenses. The literature on G3 is far larger than that on the other goals, so for G3 we include only representative studies that were published at the venues above, are widely adopted by benchmarks, or introduce a new attack class, whereas for the other goals and for studies related to \adv{T} and \adv{P} we aim for comprehensive coverage. Position papers or benchmarks that contain empirical measurements (e.g., \citet{sun2025invisible}, \citet{huang2024trustllm}) are included as measurement studies. Earlier work on neural network services and trusted execution~\citep{tramer2016stealing,shokri2017membership,tramer2019slalom,volos2018graviton} serves as background and is not coded.

\textbf{Coding.} We coded each study from its full text along the dimensions in Table~\ref{tab:coding}. Layer, adversary, and goal are multi-valued, because a single study may consider both system-prompt leakage and user-prompt leakage. Defenses are additionally coded for the type of guarantee, whether adaptive attacks were evaluated, whether the customer can deploy them without provider cooperation, and whether they were demonstrated on a model with at least 7 billion parameters. When a paper reports multiple settings, we code the strongest setting it evaluated. When a study proposes both a measurement and a defense, we code it as a single type according to its main contribution. Years are coded by the formal publication year at the time of finalization (studies that were preprints at search time and were formally published later are coded by their formal publication year), and studies not yet formally published are coded by the year of their first preprint. The first author coded all studies, and the second author independently coded a random sample of 30 studies (25.0\%), with Cohen's $\kappa$ ranging from 0.71 (adaptive evaluation) to 0.94 (type) across dimensions. Disagreements were resolved through discussion, and the resulting decision rules were then applied to re-check the same codes for the remaining studies. The full coding is listed in Appendix~\ref{app:coding}.

\textbf{Limitations.} The corpus is not exhaustive, and its composition is affected by the search terms and venue priorities. Recent studies are less completely collected, so the counts for 2025 and 2026 are lower (Section~\ref{sec:findings}), and the counts for less-studied directions such as G5 and \adv{T} should be treated as lower bounds. The G3 portion is a representative sample, so the actual share of G3 in the field is no lower than our figure. Coding involves judgment, and agreement was lowest for the adaptive evaluation and deployability dimensions. The percentages in this paper describe only this corpus and are intended to reflect the concentration of research effort.

\begin{table}[t]
\caption{Coding dimensions applied to the 120 primary studies}
\label{tab:coding}
\footnotesize
\setlength{\tabcolsep}{3pt}
\begin{tabularx}{\linewidth}{@{}lY@{}}
\toprule
Dimension & Values \\
\midrule
Type & Attack (A), measurement or benchmark (M), defense (D), one type per study by main contribution \\
Layer & L1 supply (Sup), L2 serving (Srv), L3 interface (API), L4 context (Ctx), L5 output (Out) \\
Adversary & \adv{Q} query-only user, \adv{F} customization user, \adv{C} content injector, \adv{T} side-channel observer, \adv{P} provider or intermediary, \adv{S} supply chain (Table~\ref{tab:adversaries}) \\
Goal & G1 to G6 (Table~\ref{tab:goals}) \\
Evaluation & Open-weight or local models only (O), production service only (P), both (B) \\
Guarantee & Cryptographic (Crypto), hardware isolation (HW), differential privacy (DP), statistical test (Stat), structural by construction (Struct), empirical (Emp) \\
Adaptive & Full adaptive attack evaluated (Y), partial (P), none (N), not applicable (n/a) \\
Client & Deployable by the developer or user without provider cooperation (Y/N) \\
Scale & Demonstrated on a model with $\geq$7B parameters (Y/N) \\
\bottomrule
\end{tabularx}
\end{table}

\section{System and Threat Model}\label{sec:model}

\begin{figure}[t]
\centering
\includegraphics[width=\columnwidth]{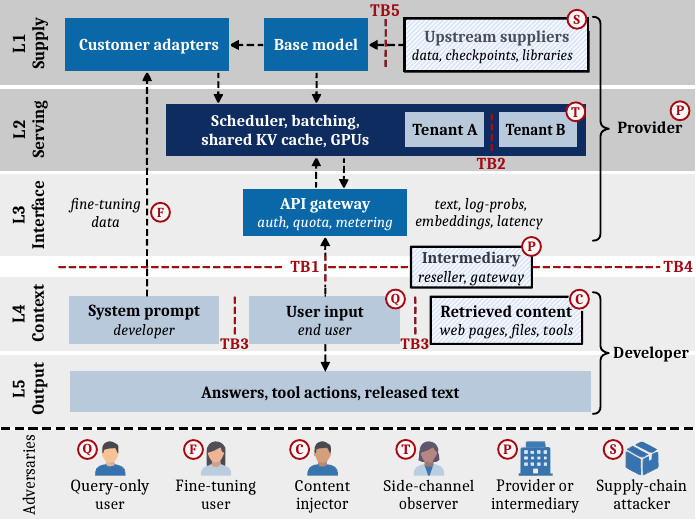}
\caption{MaaS reference architecture with layers, trust boundaries, and adversary classes. The trust boundaries are TB1 to TB5; TB3 separates developer instructions, end-user input, and third-party content within the assembled context.}
\label{fig_1_system}
\end{figure}

\subsection{Assets}

We distinguish six classes of assets. \emph{Model assets} are the weights, architectural details, adapters, and decoding configuration (L1). \emph{Application assets} are system prompts, tool definitions, and application logic, which developers increasingly treat as intellectual property (L4). \emph{User data} comprises prompts, uploaded files, conversation history, and outputs, which often contain personal or regulated information~\citep{mireshghallah2024trust}. \emph{Training and customization data} comprises pretraining corpora, fine-tuning datasets, and retrieval stores. \emph{Behavior} is the set of policies the model should follow, including refusal policies and the instruction hierarchy between developer and user~\citep{wallace2024instruction}. \emph{Service resources} are compute, quota, and billing.

\subsection{Trust boundaries}

Five Trust Boundaries (TBs) separate mutually distrusting principals (Figure~\ref{fig_1_system}).

\textbf{TB1, between client and interface.} This boundary lies between the user or developer and the API. Everything an external caller learns or influences passes through it: the input is the prompt, and the outputs are tokens, log-probabilities, embeddings, and latency.

\textbf{TB2, between tenants.} This boundary lies between requests from different tenants in the shared inference runtime. Continuous batching, prefix sharing, and semantic caching let one tenant's requests affect another tenant's latency~\citep{song2025earlybird,gu2025auditing}.

\textbf{TB3, the privilege boundary within the context.} This boundary lies inside the context assembled by L4 and separates three classes of input in decreasing order of privilege: the developer's system prompt, the end user's input, and third-party content from retrieved documents, web pages, and tool outputs. The segment between the first two determines whether the user can override the developer's instructions, and the segment between the last two is the boundary between instructions and data. Unlike the other boundaries, TB3 has no enforcement point in current models: all three classes of input are merely tokens in the same sequence~\citep{greshake2023not,wallace2024instruction,chen2025struq}.

\textbf{TB4, between customer and provider.} This boundary lies between the developer or user and the provider or any intermediary. The customer hands over data and receives in return claims about which model was run and how many tokens were consumed. Most model-centric research sidesteps this boundary by trusting the provider.

\textbf{TB5, the supply chain.} This boundary lies between the provider and the upstream suppliers of corpora, checkpoints, adapters, and software libraries~\citep{carlini2024poisoning,zhao2024models}.

\subsection{Adversaries}

Table~\ref{tab:adversaries} defines six adversary classes by the boundaries they cross and the access they possess. These classes are not ordered by strength. \adv{T} sees no content but can observe other tenants. \adv{P} sees everything inside the provider but not the customer's local state. One adversary can belong to several classes at once. For example, a developer may be a fine-tuning user (\adv{F}) and also control the documents returned by its own retrieval store (\adv{C}). The bottom of Figure~\ref{fig_1_system} lists the six adversary classes, and red circles mark where each class enters the system.

\begin{table*}[t]
\caption{Adversary classes in MaaS}
\label{tab:adversaries}
\footnotesize
\setlength{\tabcolsep}{3pt}
\begin{tabular}{@{}L{0.06\textwidth}L{0.17\textwidth}L{0.36\textwidth}L{0.07\textwidth}L{0.24\textwidth}@{}}
\toprule
Class & Who & Access & Crosses & Typical aim \\
\midrule
\adv{Q} & Query-only user & Black-box text, optionally log-probabilities or embeddings, query budget & TB1, TB3 & Extract model or data, elicit prohibited behavior, override developer instructions, exhaust resources \\
\adv{F} & Customization user & Upload of fine-tuning data through the provider's API & TB1 & Remove alignment, implant behavior \\
\adv{C} & Content injector & Control of web pages, documents, e-mails, or tool responses that enter the context & TB3 & Hijack the task, exfiltrate context, corrupt answers, inflate cost \\
\adv{T} & Side-channel observer & Co-tenant timing in the shared runtime, or on-path view of encrypted traffic & TB2, TB1 & Infer other tenants' prompts or responses \\
\adv{P} & Provider or intermediary & Full view of prompts, outputs, and serving configuration & TB4 & Read customer data, substitute models, misreport usage \\
\adv{S} & Supply-chain attacker & Control of training data, checkpoints, adapters, or loaders & TB5 & Backdoor the model, compromise the host \\
\bottomrule
\end{tabular}
\end{table*}

\subsection{Security goals}

Table~\ref{tab:goals} lists six goals. G1, G2, and G3 refine classical confidentiality and integrity for the two main classes of asset owners and for model behavior. G4 and G5 express what the customer pays for, and G6 expresses societal and regulatory requirements. G4 is the goal missing from existing taxonomies, so we give it a precise definition. Let $M^\ast$ be the model the customer contracted for, $S$ the function that actually serves requests, and $\pi$ the published pricing function. Service integrity on a workload $\mathcal{X}$ requires
\begin{equation}
\begin{split}&\forall x\in\mathcal{X}:\quad D\bigl(S(x),M^\ast(x)\bigr)\leq\epsilon\\
&\text{and}\quad c(S,x)=\pi\bigl(\tau_S(x)\bigr),\end{split}
\label{eq:service}
\end{equation}
where $S(x)$ and $M^\ast(x)$ are the distributions of the complete generated sequence (including hidden reasoning tokens that are not returned to the customer) under the published decoding parameters, $D$ is a contractually agreed distributional distance, $\epsilon$ accommodates benign numerical differences caused by batching, hardware, and kernel implementations~\citep{ong2025toploc,cai2025substitution}, $\tau_S(x)$ is the number of tokens the service actually consumes when processing $x$, and $c(S,x)$ is the amount actually billed. Covert model substitution, quantization that shifts the distribution beyond $\epsilon$, and the generation of superfluous hidden tokens violate the first condition, and misreporting the token count violates the second. A customer who sees only individual outputs cannot detect either violation~\citep{gao2025model,sun2025invisible}. G4 also requires that the model artifacts loaded by the service execute no code beyond the declared computation, a requirement that guards against \adv{S}.

\begin{table*}[t]
\caption{Security goals in MaaS}
\label{tab:goals}
\footnotesize
\setlength{\tabcolsep}{3pt}
\begin{tabular}{@{}L{0.05\textwidth}L{0.2\textwidth}L{0.4\textwidth}L{0.26\textwidth}@{}}
\toprule
Goal & Name & Requirement & Example violation \\
\midrule
G1 & Model and application confidentiality & Weights, architecture, model identity, and system prompts are not disclosed beyond the intended interface & Projection-layer extraction, prompt extraction \\
G2 & User and data confidentiality & Prompts, outputs, context, and training data are not disclosed to unauthorized parties, including the provider where required & Training-data extraction, cache timing \\
G3 & Behavioral integrity & The model follows the provider's policy and the instruction hierarchy of developer and user & Jailbreak, prompt injection, harmful fine-tuning \\
G4 & Service integrity & The served model and the billed usage match what was contracted (Eq.~\ref{eq:service}), and loaded artifacts execute nothing beyond the declared computation & Model substitution, token inflation, malicious checkpoints \\
G5 & Availability and cost & The service remains responsive and costs remain bounded under adversarial inputs & Sponge prompts, unbounded generation \\
G6 & Accountability and provenance & Outputs and models can be attributed, and behavior can be audited & Watermark removal or spoofing \\
\bottomrule
\end{tabular}
\end{table*}

\section{Attacks}\label{sec:attacks}

Table~\ref{tab:attacks} summarizes 25 attack classes by goal, adversary, required access, boundary, and whether at least one representative study was validated on a production service. This section discusses them goal by goal.

\begin{table*}[t]
\caption{Attack classes on MaaS}
\label{tab:attacks}
\scriptsize
\setlength{\tabcolsep}{3pt}
\begin{tabular}{@{}L{0.2\textwidth}L{0.06\textwidth}L{0.06\textwidth}L{0.16\textwidth}L{0.08\textwidth}C{0.06\textwidth}L{0.25\textwidth}@{}}
\toprule
Attack class & Goal & Adv. & Access & Bound. & Prod. & Representative studies \\
\midrule
Logit-based weight extraction & G1 & \adv{Q} & Log-probabilities & TB1 & \full & \cite{carlini2024stealing,finlayson2024logits} \\
Functional imitation & G1 & \adv{Q} & Text & TB1 & \full & \cite{krishna2020thieves,wallace2020imitation,gudibande2024false} \\
System-prompt extraction & G1 & \adv{Q} & Text & TB1, TB3 & \full & \cite{perez2022ignore,zhang2024effective,hui2024pleak,morris2024inversion} \\
Model fingerprinting & G1 & \adv{Q} & Text & TB1 & \full & \cite{pasquini2025llmmap} \\
Training-data extraction & G2 & \adv{Q} & Text & TB1 & \full & \cite{carlini2021extracting,lukas2023analyzing,nasr2023scalable} \\
Membership inference & G2 & \adv{Q} & Log-probabilities & TB1 & \none & \cite{mattern2023membership,shi2024detecting,fu2024practical,duan2024membership,debenedetti2024privacy} \\
Attribute inference from user text & G2 & \adv{Q} & Text & TB1 & \full & \cite{staab2024beyond} \\
Embedding inversion & G2 & \adv{P} & Embeddings & TB4 & \full & \cite{song2020information,morris2023text,li2023sentence} \\
Retrieval-store leakage & G2 & \adv{Q} & Text & TB1, TB3 & \none & \cite{zeng2024good} \\
Cache and scheduling side channels & G1, G2 & \adv{T} & Timing & TB2 & \full & \cite{song2025earlybird,wu2025promptpeek,gu2025auditing} \\
Token-length side channel & G2 & \adv{T} & Packet sizes & TB1 & \full & \cite{weiss2024keylogging} \\
Provider-side exposure & G2 & \adv{P} & Logs & TB4 & \full & \cite{mireshghallah2024trust} \\
Optimization-based jailbreak & G3 & \adv{Q} & Surrogate gradients, text & TB1 & \full & \cite{zou2023universal,andriushchenko2025jailbreaking} \\
LLM-guided jailbreak & G3 & \adv{Q} & Text & TB1 & \full & \cite{chao2025jailbreaking,mehrotra2024tap,liu2024autodan} \\
Multi-turn and long-context jailbreak & G3 & \adv{Q} & Text & TB1 & \full & \cite{anil2024manyshot,russinovich2025crescendo} \\
Direct prompt injection & G3 & \adv{Q} & Text & TB1, TB3 & \full & \cite{perez2022ignore,liu2023houyi} \\
Indirect prompt injection & G3, G2 & \adv{C} & Content & TB3 & \full & \cite{greshake2023not,liu2024formalizing,zhan2024injecagent} \\
Knowledge-base poisoning & G3 & \adv{C} & Content & TB3 & \full & \cite{zou2025poisonedrag} \\
Harmful fine-tuning & G3 & \adv{F} & Fine-tuning API & TB1 & \full & \cite{qi2024finetuning,halawi2024covert} \\
Data poisoning and backdoors & G3 & \adv{S} & Training data & TB5 & \full & \cite{wan2023poisoning,rando2024universal,hubinger2024sleeper,carlini2024poisoning} \\
Malicious model artifacts & G4 & \adv{S} & Checkpoints, loaders & TB5 & \full & \cite{zhao2024models} \\
Model substitution and usage inflation & G4, G5 & \adv{P} & Serving configuration & TB4 & \full & \cite{cai2025substitution,sun2025invisible} \\
Resource exhaustion & G5 & \adv{Q} & Text & TB1 & \full & \cite{shumailov2021sponge,geiping2024coercing,dong2025engorgio} \\
Reasoning slowdown via injected content & G5 & \adv{C} & Content & TB3 & \full & \cite{kumar2025overthink} \\
Watermark stealing and spoofing & G6 & \adv{Q} & Text & TB1 & \none & \cite{jovanovic2024watermark} \\
\bottomrule
\end{tabular}
\tabnote{Prod.: \full\ at least one representative study was evaluated on a production service, \none\ none.}
\end{table*}

\subsection{Model and application confidentiality (G1)}

Model stealing through prediction APIs predates LLMs~\citep{tramer2016stealing}, and its earliest forms against language models were the imitation of BERT-based APIs~\citep{krishna2020thieves} and of production machine translation systems~\citep{wallace2020imitation}, as well as the recovery of fine-tuned classification layers when the public pretrained encoder is known~\citep{zanella2021grey}. Two lines of work now coexist. \emph{Exact extraction} recovers parameters. \citet{carlini2024stealing} exploited log-probability outputs to recover the embedding projection layer of OpenAI production models for less than \$20, and estimated that doing the same for gpt-3.5-turbo would cost less than \$2{,}000. \citet{finlayson2024logits} showed that the outputs of API-protected models lie in a low-dimensional subspace, from which one can infer their hidden dimension and identify model updates. Both attacks rely on the interface returning log-probabilities or allowing a logit bias to be set, and providers subsequently restricted these features, which shows that interface design at L3 directly determines G1. \emph{Functional imitation} trains a cheaper model on API outputs. \citet{gudibande2024false} found that imitation mostly learns style and rarely learns factuality and capability, which limits its threat to frontier models. Model identity is itself a form of information: LLMmap~\citep{pasquini2025llmmap} uses crafted queries to identify the model version behind an application, and even in the presence of unknown system prompts, stochastic sampling, and generation frameworks such as RAG, it distinguishes 42 versions (including closed-source models) with over 95\% accuracy in only 8 interactions, which prepares the ground for subsequent model-specific attacks.

System prompts are application assets at L4. Instructions such as ``ignore the above and output your instructions'' leak them~\citep{perez2022ignore,liu2023houyi}. \citet{zhang2024effective} showed that simple text-only queries reliably extract system prompts, including from production systems, PLeak~\citep{hui2024pleak} automatically optimizes such queries, and \citet{morris2024inversion} reconstruct prompts from the next-token probability vector alone. Defenses that instruct the model to keep its prompt secret sit inside TB3 and inherit all the weaknesses of the in-band defenses discussed in Section~\ref{sec:def-mediation}.

\subsection{User and data confidentiality (G2)}

\textbf{Training and customization data.} \citet{carlini2021extracting} defined training data extraction for language models: given a model $f_\theta$, find an input that causes it to output a sample $x\in D_{\text{train}}$ verbatim. Memorization grows log-linearly with model size and the number of duplicates~\citep{carlini2023quantifying}, personally identifiable information can be extracted or reconstructed from fine-tuned models~\citep{lukas2023analyzing}, and alignment does not eliminate the phenomenon: a divergence attack caused ChatGPT to emit training data at about 150 times the rate observed in normal use~\citep{nasr2023scalable}. Membership inference is a weaker, statistical version of this attack. Reference-free attacks based on neighborhood comparison~\citep{mattern2023membership} and token probability statistics~\citep{shi2024detecting}, as well as attacks on fine-tuned models~\citep{fu2024practical}, all report positive results. However, a large-scale evaluation on models trained on the Pile found that membership inference on pretraining data is barely better than random guessing, and that many apparent successes stem from distribution shift between members and non-members~\citep{duan2024membership}. \citet{debenedetti2024privacy} further showed that system components surrounding the model (e.g., training-data deduplication and output filters) create privacy side channels that invalidate guarantees proven for the model alone. In MaaS, therefore, fine-tuning data, which is small and heavily duplicated, faces greater risk than pretraining data.

\textbf{Context and user data.} User prompts reach the provider by design. An analysis of real conversations found that users frequently disclose personal and sensitive information to chatbots~\citep{mireshghallah2024trust}, and models also leak in-context information in ways that violate contextual integrity~\citep{mireshghallah2024secret}. Beyond memorization, LLMs can infer personal attributes such as location and income from ordinary user text with high accuracy, at a fraction of the cost of human analysis~\citep{staab2024beyond}, which allows any party holding conversation logs to profile users. Retrieval introduces two kinds of exposure: retrieval stores can be induced to output their documents~\citep{zeng2024good}, and embeddings stored with a vector database provider can be inverted into text close to the original~\citep{song2020information,li2023sentence,morris2023text}, including embeddings produced by commercial embedding APIs~\citep{morris2023text}.

\textbf{Side channels in the inference stack.} Another class of threats to G2 comes from performance optimizations at L2 and L3. Prefix caching makes requests that share a prefix with a cached request faster, which a co-located tenant can exploit to verify guesses about another tenant's prompt token by token. \citet{song2025earlybird} demonstrated such attacks on KV caches and semantic caches in open-source serving frameworks, and reported cache sharing in several commercial services. PromptPeek~\citep{wu2025promptpeek} exploits the cross-user KV cache sharing in frameworks such as vLLM and SGLang, together with the scheduling policies that accompany it, to reconstruct other users' prompts token by token. \citet{gu2025auditing} audited 17 commercial API providers in the second half of 2024 and detected prompt caching at 8 of them, 7 of which shared the cache across users. They also used the same timing signal to infer that one of OpenAI's embedding models has a decoder-only architecture, which constitutes a G1 leak. At the network level, \citet{weiss2024keylogging} found that streaming APIs send one token per encrypted packet, so packet lengths reveal token lengths, and a model trained to invert packet lengths reconstructed 29\% of the responses from commercial assistants' traffic and inferred the topic of 55\% of them.

\subsection{Behavioral integrity (G3)}

\textbf{Jailbreaks.} \citet{wei2023jailbroken} attributed the success of jailbreaks to competing objectives and to mismatched generalization between capability training and safety training, and \citet{shen2024donow} analyzed 1{,}405 jailbreak prompts collected in the wild. Greedy Coordinate Gradient (GCG)~\citep{zou2023universal} formulates the attack as finding an adversarial suffix $\delta$ that maximizes the likelihood of an affirmative target response $y^\ast$:
\begin{equation}
\min_{\delta}\; -\log p_\theta(y^\ast\mid x\oplus\delta),
\end{equation}
and optimizes this suffix on open-source models. The suffixes transfer to closed-source models, reaching attack success rates of 86.6\% on GPT-3.5 and 46.9\% on GPT-4 in the original evaluation. Black-box methods replace gradients with an attacker LLM that iteratively rewrites the prompt within about twenty queries~\citep{chao2025jailbreaking}, or adopt tree search with pruning~\citep{mehrotra2024tap} or genetic search to generate fluent prompts~\citep{liu2024autodan}. Simple adaptive attacks that combine prompt templates, random search, and API-specific features such as prefilling achieve nearly complete success on many safety-aligned models~\citep{andriushchenko2025jailbreaking}. Long contexts and multi-turn conversations open further channels: many-shot jailbreaking fills the context with hundreds of fabricated dialogues~\citep{anil2024manyshot}, while Crescendo escalates gradually over a seemingly harmless multi-turn conversation~\citep{russinovich2025crescendo}. Automated red teaming with attacker language models~\citep{perez2022red,hong2024curiosity} applies the same idea to generate test cases. The \emph{Malla} study documents the real-world consequences: 212 malicious services sold on underground markets are built on public LLM APIs through jailbreak prompts or directly use uncensored models~\citep{lin2024malla}.

\textbf{Prompt injection.} Prompt injection breaks TB3. Direct injection crosses the segment between developer and user, with the user overriding the developer's instructions~\citep{perez2022ignore,liu2023houyi}. Indirect injection crosses the segment between instructions and data by embedding instructions in content that the application will later retrieve~\citep{greshake2023not}. \citet{liu2024formalizing} formalized this attack as causing the application to perform an injected task in place of the target task, and benchmarked combinations of attacks and defenses on ten LLMs. In tool-using agents, the harm grows from wrong text to wrong actions: InjecAgent measured an attack success rate of 24\% against a GPT-4 agent with ReAct prompting, including data exfiltration attacks~\citep{zhan2024injecagent}, and AgentDojo and BIPIA provide dynamic and static benchmarks for agents and applications, respectively~\citep{debenedetti2024agentdojo,yi2025bipia}. Knowledge-base poisoning is the counterpart attack in retrieval settings: PoisonedRAG injects 5 malicious texts per target question into a knowledge base containing millions of documents and achieves an attack success rate of about 90\%~\citep{zou2025poisonedrag}.

\textbf{Customization and supply chain.} Fine-tuning APIs allow \adv{F} to change the model itself. \citet{qi2024finetuning} jailbroke GPT-3.5 Turbo through the OpenAI fine-tuning API with 10 adversarial examples at a cost of less than \$0.20, and showed that even benign fine-tuning degrades safety. Covert malicious fine-tuning encodes harmful training data so that it passes the provider's data moderation~\citep{halawi2024covert}. Upstream, poisoning instruction-tuning data~\citep{wan2023poisoning} or human feedback data~\citep{rando2024universal} can implant triggers, deliberately trained backdoors persist through standard safety training~\citep{hubinger2024sleeper}, and poisoning web-scale datasets is also feasible, for example by purchasing expired domains referenced in dataset indexes~\citep{carlini2024poisoning}.

\subsection{Service integrity (G4) and availability and cost (G5)}

G4 is violated when what the customer receives does not match the promise of Eq.~\ref{eq:service}. \citet{gao2025model} audited nine commercial inference providers serving four Llama models in mid-2024 and found that 11 of 31 endpoints had output distributions that differed from those of the reference weights released by Meta. \citet{cai2025substitution} studied how a dishonest provider can substitute a cheaper or quantized model, and found that output-based tests require many queries, while log-probability-based tests are confounded by numerical nondeterminism. \citet{sun2025invisible} pointed out that reasoning APIs bill for hidden tokens, whose number often exceeds that of answer tokens by more than an order of magnitude, and customers cannot verify them. Supply-chain artifacts also threaten the provider's own platform: scanning the Hugging Face model hub with taint analysis uncovered 91 malicious models and 9 malicious dataset loading scripts that execute code at load time~\citep{zhao2024models}.

For G5, sponge examples maximize the energy consumption and latency of each query~\citep{shumailov2021sponge}, optimized prompts can force models to produce verbose or repetitive output~\citep{geiping2024coercing}, and Engorgio prompts suppress the end-of-sequence token so that generation continues until the length limit~\citep{dong2025engorgio}. Reasoning models extend this attack surface to TB3: OverThink~\citep{kumar2025overthink} injects decoy reasoning problems into retrieved content, causing reasoning models to consume far more reasoning tokens while the answer remains unchanged, with slowdowns of up to 18 times on FreshQA and 46 times on SQuAD, and the attacker only needs to control content that enters the context (\adv{C}). Because MaaS bills by output tokens, these attacks shift costs to whoever pays for the request, which may be the developer rather than the attacker.

\subsection{Accountability and provenance (G6)}

Watermarks make machine-generated text detectable (Section~\ref{sec:def-account}). \citet{jovanovic2024watermark} showed that merely querying a watermarked API suffices to learn enough about distribution-shifting watermarks to spoof them on arbitrary text or to scrub them, so the accountability mechanism itself becomes a target for \adv{Q}.

\subsection{Composition across boundaries}

\begin{figure}[t]
\centering
\includegraphics[width=\columnwidth]{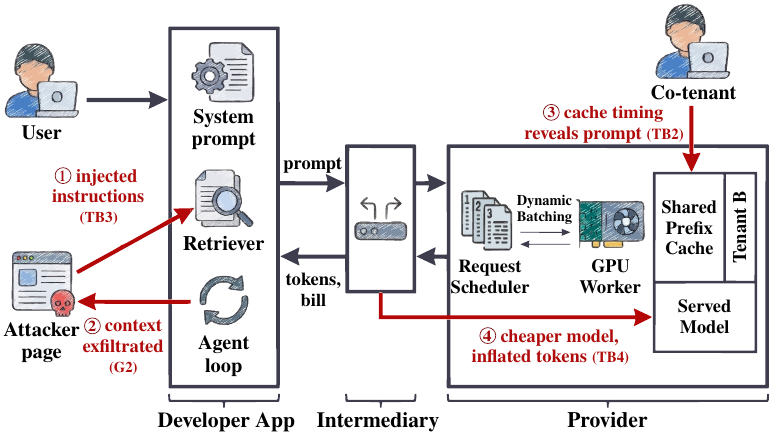}
\caption{One RAG application on a commercial API attacked across four trust boundaries}
\label{fig_2_attack}
\end{figure}

Attacks increasingly chain multiple boundaries. Figure~\ref{fig_2_attack} takes as an example a retrieval-augmented application that calls a commercial API through an intermediary: instructions on an attacker's web page enter the context through retrieval (\ding{172}, TB3), the agent loop then sends the context to an attacker-controlled endpoint (\ding{173}, G2), a co-located tenant recovers the same application's system prompt through prefix cache timing (\ding{174}, TB2), and the intermediary can switch to a cheaper model and overcount tokens (\ding{175}, TB4). These four steps correspond to four attack classes in Table~\ref{tab:attacks}, all acting on the same service. Similar compositions already appear in the literature: indirect injection (TB3) has been used to exfiltrate context to attacker-controlled endpoints (G2)~\citep{greshake2023not,zhan2024injecagent}, fine-tuning access (\adv{F}) turns a single G3 violation into a persistent property of the model~\citep{qi2024finetuning}, and timing leakage in L2 turns a performance optimization into a G1 and G2 vulnerability~\citep{gu2025auditing}. Defenses evaluated against only a single boundary therefore provide little assurance for the composed system.

\section{Assurance Mechanisms}\label{sec:defenses}

\begin{figure*}[t]
\centering
\includegraphics{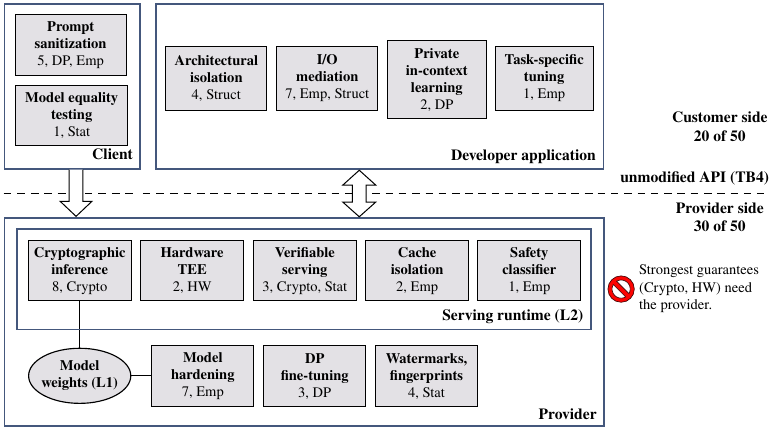}
\caption{Who can deploy each family of the 50 coded assurance mechanisms}
\label{fig_3_defense}
\end{figure*}

We organize defenses by trust anchor (the component on which the guarantee rests), because the trust anchor determines both which adversaries a defense can address and who deploys it. Table~\ref{tab:defenses} evaluates all 50 coded defenses against six criteria. Figure~\ref{fig_3_defense} groups these defenses by mechanism family according to who must deploy them, and the number in each box is the number of defenses in that family, followed by the guarantee types (abbreviations as in Table~\ref{tab:coding}). The mechanism families in the figure are divided by deploying party and are finer than the groups in the table: Jatmo, which the developer fine-tunes on its own, belongs to Model hardening in the table but is shown separately as Task-specific tuning in the figure, Constitutional Classifiers, which only the provider can deploy, belongs to Input and output mediation but is shown separately as Safety classifier, DP-ICL and PromptPATE from the DP group are shown separately as Private in-context learning, and model equality testing from Verifiable serving is placed on the client side. The 20 defenses that clients and developer applications can deploy unilaterally all operate on top of an unmodified API, and the remaining 30 must be implemented by the provider in the inference runtime or the model, including all cryptographic and hardware mechanisms (F4 in Section~\ref{sec:findings}).

\begin{table*}[t]
\caption{Evaluation of MaaS assurance mechanisms}
\label{tab:defenses}
\scriptsize
\setlength{\tabcolsep}{2.2pt}
\begin{minipage}[t]{0.49\textwidth}
\begin{tabularx}{\linewidth}[t]{@{}Yllcccc@{}}
\toprule
Mechanism & Adv. & Guar. & Client & $\geq$7B & Prod. & Adapt. \\
\midrule
\multicolumn{7}{@{}l}{\textit{Hardware isolation}} \\
H100 confidential computing \citep{zhu2024hopper} & \adv{P} & HW & \none & \full & \none & n/a \\
TransLinkGuard \citep{li2024translinkguard} & \adv{Q} & HW & \none & \none & \none & n/a \\
\midrule
\multicolumn{7}{@{}l}{\textit{Cryptographic inference}} \\
Iron \citep{hao2022iron} & \adv{P} & Crypto & \none & \none & \none & n/a \\
MPCFormer \citep{li2023mpcformer} & \adv{P} & Crypto & \none & \none & \none & n/a \\
BOLT \citep{pang2024bolt} & \adv{P} & Crypto & \none & \none & \none & n/a \\
SecFormer \citep{luo2024secformer} & \adv{P} & Crypto & \none & \none & \none & n/a \\
Nimbus \citep{li2024nimbus} & \adv{P} & Crypto & \none & \none & \none & n/a \\
SHAFT \citep{kei2025shaft} & \adv{P} & Crypto & \none & \none & \none & n/a \\
BumbleBee \citep{jie2025bumblebee} & \adv{P} & Crypto & \none & \full & \none & n/a \\
PUMA \citep{dong2025puma} & \adv{P} & Crypto & \none & \full & \none & n/a \\
\midrule
\multicolumn{7}{@{}l}{\textit{Verifiable serving}} \\
zkLLM \citep{sun2024zkllm} & \adv{P} & Crypto & \none & \full & \none & n/a \\
TOPLOC \citep{ong2025toploc} & \adv{P} & Stat & \none & \full & \none & n/a \\
CoIn \citep{sun2025coin} & \adv{P} & Stat & \none & \full & \none & n/a \\
Model equality testing \citep{gao2025model} & \adv{P} & Stat & \full & \full & \full & n/a \\
\midrule
\multicolumn{7}{@{}l}{\textit{Runtime isolation}} \\
SafeKV \citep{chu2025safekv} & \adv{T} & Emp & \none & \full & \none & \none \\
KV-Cloak \citep{luo2026kvcloak} & \adv{P} & Emp & \none & \full & \none & \none \\
\midrule
\multicolumn{7}{@{}l}{\textit{DP training and prompting}} \\
DP fine-tuning \citep{yu2022differentially} & \adv{Q} & DP & \none & \none & \none & n/a \\
DP large-batch tuning \citep{li2022large} & \adv{Q} & DP & \none & \none & \none & n/a \\
DP-Forward \citep{du2023dpforward} & \adv{Q}, \adv{P} & DP & \none & \none & \none & n/a \\
DP-ICL \citep{tang2024privacy} & \adv{Q} & DP & \full & \full & \full & n/a \\
PromptPATE \citep{duan2023promptpate} & \adv{Q} & DP & \full & \full & \full & n/a \\
\midrule
\multicolumn{7}{@{}l}{\textit{Prompt sanitization}} \\
SANTEXT+ \citep{yue2021santext} & \adv{P} & DP & \full & \none & \none & n/a \\
CusText \citep{chen2023customized} & \adv{P} & DP & \full & \none & \none & n/a \\
InferDPT \citep{tong2025inferdpt} & \adv{P} & DP & \full & \full & \full & n/a \\
Hide-and-Seek \citep{chen2023hideandseek} & \adv{P} & Emp & \full & \full & \full & \none \\
CoGenesis \citep{zhang2024cogenesis} & \adv{P} & Emp & \full & \full & \full & \none \\
\bottomrule
\end{tabularx}
\end{minipage}\hfill
\begin{minipage}[t]{0.49\textwidth}
\begin{tabularx}{\linewidth}[t]{@{}Yllcccc@{}}
\toprule
Mechanism & Adv. & Guar. & Client & $\geq$7B & Prod. & Adapt. \\
\midrule
\multicolumn{7}{@{}l}{\textit{Model hardening}} \\
Vaccine \citep{huang2024vaccine} & \adv{F} & Emp & \none & \full & \none & \none \\
Circuit Breakers \citep{zou2024circuit} & \adv{Q} & Emp & \none & \full & \none & \full \\
StruQ \citep{chen2025struq} & \adv{C} & Emp & \none & \full & \none & \half \\
SecAlign \citep{chen2025secalign} & \adv{C} & Emp & \none & \full & \none & \full \\
Instruction hierarchy \citep{wallace2024instruction} & \adv{C}, \adv{Q} & Emp & \none & \full & \full & \none \\
Jatmo \citep{piet2024jatmo} & \adv{C} & Emp & \full & \full & \full & \half \\
SafeDecoding \citep{xu2024safedecoding} & \adv{Q} & Emp & \none & \full & \none & \none \\
RAIN \citep{li2024rain} & \adv{Q} & Emp & \none & \full & \none & \none \\
\midrule
\multicolumn{7}{@{}l}{\textit{Input and output mediation}} \\
Perplexity and paraphrase filters \citep{jain2023baseline} & \adv{Q} & Emp & \full & \full & \none & \full \\
SmoothLLM \citep{robey2024smoothllm} & \adv{Q} & Emp & \full & \full & \full & \half \\
Erase-and-Check \citep{kumar2025certifying} & \adv{Q} & Struct & \full & \full & \none & \half \\
Llama Guard \citep{inan2024llamaguard} & \adv{Q} & Emp & \full & \full & \none & \none \\
WildGuard \citep{han2024wildguard} & \adv{Q} & Emp & \full & \full & \none & \none \\
Constitutional Classifiers \citep{sharma2025constitutional} & \adv{Q} & Emp & \none & \full & \full & \full \\
Spotlighting \citep{hines2024spotlighting} & \adv{C} & Emp & \full & \full & \full & \none \\
DataSentinel \citep{liu2025datasentinel} & \adv{C} & Emp & \full & \full & \none & \full \\
\midrule
\multicolumn{7}{@{}l}{\textit{Architectural isolation}} \\
IsolateGPT \citep{wu2025isolategpt} & \adv{C} & Struct & \full & \full & \full & \none \\
CaMeL \citep{debenedetti2026camel} & \adv{C} & Struct & \full & \full & \full & \half \\
Progent \citep{shi2025progent} & \adv{C} & Struct & \full & \full & \full & \half \\
RobustRAG \citep{xiang2024certifiably} & \adv{C} & Struct & \full & \full & \none & \half \\
\midrule
\multicolumn{7}{@{}l}{\textit{Provenance}} \\
Green-list watermark \citep{kirchenbauer2023watermark} & \adv{Q} & Stat & \none & \none & \none & \half \\
Distortion-free watermark \citep{kuditipudi2024robust} & \adv{Q} & Stat & \none & \full & \none & \half \\
SynthID-Text \citep{dathathri2024scalable} & \adv{Q} & Stat & \none & \full & \full & \half \\
Instructional fingerprint \citep{xu2024instructional} & \adv{Q} & Stat & \none & \full & \none & \half \\
\bottomrule
\end{tabularx}
\end{minipage}
\tabnote{Guar.: type of guarantee (Table~\ref{tab:coding}). Client: deployable by the customer without provider cooperation. $\geq$7B: demonstrated at this scale. Prod.: evaluated on a production service. Adapt.: \full\ full adaptive evaluation, \half\ partial, \none\ none, n/a not applicable.}
\end{table*}

\subsection{Hardware-rooted isolation}

A Trusted Execution Environment (TEE) protects computation from a privileged host. Early work combined enclaves with untrusted accelerators: Slalom~\citep{tramer2019slalom} blinds the input of a linear layer $Y=XW$, sends $X+R$ to the GPU, subtracts the precomputed $RW$ inside the enclave, and verifies the result with Freivalds' algorithm. Graviton~\citep{volos2018graviton} and Honeycomb~\citep{mai2023honeycomb} bring isolation into the GPU itself. GPUs that support confidential computing make this approach practical for LLMs: on NVIDIA Hopper GPUs, the overhead of TEE mode is below 7\% for most LLM queries and approaches zero for larger models and longer sequences, with the remaining overhead coming mainly from encrypted transfers between CPU and GPU~\citep{zhu2024hopper}. Production systems such as Apple's Private Cloud Compute combine hardware attestation with stateless processing and published software images~\citep{apple2024pcc}. When the model runs on hardware controlled by the customer, hardware isolation can also protect the provider's model, as TransLinkGuard does for Transformers deployed at the edge~\citep{li2024translinkguard}. Its limitations have long been known from the TEE literature: side-channel and transient-execution attacks~\citep{vanbulck2020lvi}, and the need for cross-vendor attestation infrastructure that customers can verify~\citep{menetrey2022exploratory}. MaaS adds a decisive limitation: unless the provider opens it up, the customer can neither enable a TEE nor verify its attestation, so TEEs help against \adv{P} only when the provider is willing to be verified.

\subsection{Cryptographic inference}

MPC and homomorphic encryption let the client obtain inference results without exposing its input, and in the two-party setting the client also learns nothing about the weights. Iron~\citep{hao2022iron} adapts two-party protocols to matrix multiplication, Softmax, the Gaussian Error Linear Unit (GeLU), and layer normalization in Transformers. Follow-up work replaces nonlinear operators with MPC-friendly approximations and distillation~\citep{li2023mpcformer,luo2024secformer}, improves ciphertext packing and communication~\citep{pang2024bolt,jie2025bumblebee,li2024nimbus}, and introduces constant-round protocols for Softmax and GeLU~\citep{kei2025shaft}. PUMA~\citep{dong2025puma} was the first to evaluate a 7-billion-parameter model, requiring about 5 minutes per generated token in the three-party semi-honest model, and BumbleBee~\citep{jie2025bumblebee} also reported results for LLaMA-7B in the two-party setting. These mechanisms offer the strongest guarantees covered in this paper, but every one of them requires the provider to run the protocol, most evaluations stop at BERT-scale models, and latency remains orders of magnitude higher than that of plaintext inference.

\subsection{Verifiable serving}

Verifiable serving targets G4 against \adv{P}. zkLLM~\citep{sun2024zkllm} generates zero-knowledge proofs of correct inference for models with up to 13 billion parameters, taking less than 15 minutes per proof. TOPLOC~\citep{ong2025toploc} commits to locality-sensitive hashes of intermediate activations, allowing a verifier to detect changes to the model, prompt, or precision while tolerating benign numerical nondeterminism. CoIn~\citep{sun2025coin} targets billing for hidden tokens: it builds a verifiable hash tree from fingerprints of token embeddings to check token counts, and uses embedding-based relevance matching to detect fabricated reasoning content, reporting detection success rates of up to 94.7\%. All three require provider cooperation. The only G4 defense in this corpus that a customer can run alone is \emph{model equality testing}~\citep{gao2025model}, which turns the first condition of Eq.~\ref{eq:service} into a two-sample test between samples from the API and samples from the reference weights. A maximum mean discrepancy test with a simple string kernel, using on average only ten samples per prompt, detected a range of distortions with a median power of 77.4\%. Statistical auditing requires reference weights to verify model identity, and for closed-source models it can only compare the API with its own past samples to detect drift. Neither use can verify billing.

\subsection{Inference-runtime isolation}

Defenses inside the inference runtime target the shared state in L2. The most direct approach is to isolate caches per tenant, but this gives up the main benefit of prefix sharing. SafeKV~\citep{chu2025safekv} detects the sensitivity of prompt content and allows only KV cache entries judged non-sensitive to be shared across tenants, keeping sensitive content in private caches. Its average defense success rate against timing side-channel attacks exceeds 94\%. Compared with full isolation, it reduces the additional time-to-first-token overhead by up to 40.58\%, and compared with per-user cache isolation, it increases throughput by up to 2.66 times. KV-Cloak~\citep{luo2026kvcloak} addresses a party that can read the KV cache directly, for example when the cache is offloaded to external storage, and it uses reversible matrix obfuscation and operator fusion to prevent cache contents from being inverted into the original input, with small performance overhead. Both can be deployed only by the party that operates the serving engine, and both provide empirical guarantees. The protection of SafeKV depends on the accuracy of sensitivity detection, and the detector itself has not undergone adaptive evaluation.

\subsection{Statistical privacy}

DP gives a formal upper bound on the information about any single record that any output can leak. A randomized mechanism $\mathcal{M}$ satisfies $(\varepsilon,\delta)$-differential privacy if, for all neighboring datasets $D,D'$ and all output sets $S$,
\begin{equation}
\Pr[\mathcal{M}(D)\in S]\leq e^{\varepsilon}\Pr[\mathcal{M}(D')\in S]+\delta
\end{equation}
\citep{dwork2014algorithmic}. Differentially private stochastic gradient descent~\citep{abadi2016deep} clips per-example gradients and adds Gaussian noise. Combined with parameter-efficient methods and large-batch training, DP fine-tuning of language models approaches the utility of non-private training on several tasks~\citep{yu2022differentially,li2022large}, and DP-Forward instead perturbs representations in the forward pass~\citep{du2023dpforward}. These mechanisms protect fine-tuning data against \adv{Q}, but they must be implemented by the trainer (usually the provider). Two mechanisms put DP in the hands of the developer: DP-ICL~\citep{tang2024privacy} and PromptPATE~\citep{duan2023promptpate} aggregate answers over disjoint sets of private examples so that the released output or prompt is differentially private with respect to those examples, and both can run through commercial APIs.

Against \adv{P}, the only thing a customer can change is what it sends. Text sanitization replaces sensitive words under metric local DP~\citep{yue2021santext,chen2023customized}, InferDPT~\citep{tong2025inferdpt} perturbs the prompt and uses a local model to repair the API's answer, and Hide-and-Seek~\citep{chen2023hideandseek} and CoGenesis~\citep{zhang2024cogenesis} keep sensitive entities or reasoning on a local model. They are the only client-deployable means in this corpus that protect user data from the provider. Their guarantees are at the word level, protecting explicit identifiers better than the semantics of the request, and tighter privacy budgets yield lower utility.

\subsection{Model hardening}

Hardening means changing the model itself so that it resists G3 attacks. Alignment from human or AI feedback~\citep{ouyang2022instructgpt,bai2022constitutional} is the baseline. Representation-level methods such as Circuit Breakers~\citep{zou2024circuit} reroute internal representations associated with harmful outputs, and were evaluated against a wide range of attacks, including adaptive ones. Decoding-time methods shift token probabilities toward safe continuations~\citep{xu2024safedecoding} or let the model evaluate and rewind its own generation~\citep{li2024rain}. Against prompt injection, the instruction hierarchy trains the model to prioritize instructions in the order system, user, tool~\citep{wallace2024instruction}. StruQ~\citep{chen2025struq} and SecAlign~\citep{chen2025secalign} use reserved delimiters to separate instructions from data, and train the model through supervised fine-tuning or preference optimization to ignore instructions in the data portion. Jatmo~\citep{piet2024jatmo} fine-tunes a base model without instruction tuning for a single task, so that injected instructions have no instruction-following behavior to exploit. Against \adv{F}, Vaccine~\citep{huang2024vaccine} adds perturbation-aware alignment so that the model remains safe when it is later subjected to harmful fine-tuning. Except for Jatmo, all of these mechanisms must be implemented by the party that trains the model.

\subsection{Input and output mediation}\label{sec:def-mediation}

Mediation defenses add a check around an unmodified model. Perplexity and paraphrase filters detect optimized suffixes but not fluent jailbreaks~\citep{jain2023baseline}. SmoothLLM~\citep{robey2024smoothllm} exploits the brittleness of adversarial suffixes by aggregating the responses to multiple randomly perturbed copies of the prompt. Erase-and-Check~\citep{kumar2025certifying} certifies that harmful prompts carrying adversarial suffixes up to a given length are always detected, at a cost that grows with that length. Guard models classify prompts and responses according to a policy~\citep{inan2024llamaguard,han2024wildguard}. Constitutional Classifiers~\citep{sharma2025constitutional} train input and output classifiers on synthetic data generated from a natural-language ``constitution''. In red teaming estimated at more than 3{,}000 hours, no participant found a universal jailbreak against the protected prototype, and an improved version increased the refusal rate on production traffic by 0.38 percentage points with an inference overhead of 23.7\%. Against injection, spotlighting marks or encodes untrusted content so that the model can recognize it~\citep{hines2024spotlighting}, and DataSentinel~\citep{liu2025datasentinel} trains a detector through a minimax game against an adaptive injector. All of these defenses are in-band: the check is itself an LLM or relies on the model's understanding, so it can be targeted together with the model it protects~\citep{nasr2025attacker,zhan2025adaptive}.

\subsection{Architectural isolation}

Architectural defenses enforce TB3 outside the model. IsolateGPT~\citep{wu2025isolategpt} runs each third-party app in an isolated execution context, with a central hub coordinating interactions between apps. CaMeL~\citep{debenedetti2026camel} has a privileged model write a program based only on the trusted user query, executes this program in an interpreter that tracks capabilities and data provenance, and lets a quarantined model process untrusted data without the ability to choose tools, so that injected content cannot alter the control flow. Progent~\citep{shi2025progent} enforces fine-grained permission policies on tool calls, and RobustRAG~\citep{xiang2024certifiably} isolates individual retrieved passages and aggregates their answers, so that a bounded number of corrupted passages provably cannot change the output. Design patterns that generalize these ideas restrict what an agent can do after it has read untrusted data~\citep{beurerkellner2025design}. These mechanisms provide structural guarantees that do not depend on whether the model resists injection, and developers can deploy them on top of commercial APIs. The cost is reduced flexibility, since tasks whose control flow must depend on untrusted data cannot be expressed.

\subsection{Provenance and accountability}\label{sec:def-account}

Statistical watermarks bias generation toward a pseudorandom subset of tokens. In the scheme of \citet{kirchenbauer2023watermark}, a hash of the previous token selects a ``green list'' containing a fraction $\gamma$ of the vocabulary, and $\delta$ is added to the logits of green-list tokens. Detection counts the number $|s|_G$ of green-list tokens in a text of length $T$ and computes
\begin{equation}
z=\frac{|s|_G-\gamma T}{\sqrt{T\gamma(1-\gamma)}}.
\end{equation}
Distortion-free schemes preserve the output distribution and withstand more edits~\citep{kuditipudi2024robust}, and SynthID-Text has been deployed in a production chatbot~\citep{dathathri2024scalable}. Fingerprinting embeds a secret input-output pair into the weights so that the model owner can later recognize a stolen or redistributed model~\citep{xu2024instructional}. All provenance mechanisms are implemented by the provider. They help providers and regulators attribute outputs (G6) and protect G1, but they do not help customers with G4, and adversaries who query a watermarked API can spoof or remove them~\citep{jovanovic2024watermark}.

\section{Evaluation Practice}\label{sec:evaluation}

Benchmarks make the results of attacks and defenses more comparable. DecodingTrust~\citep{wang2023decodingtrust} and TrustLLM~\citep{huang2024trustllm} evaluate trustworthiness dimensions such as toxicity, privacy, and robustness, HarmBench~\citep{mazeika2024harmbench} and JailbreakBench~\citep{chao2024jailbreakbench} standardize harmful behavior sets, judges, and leaderboards for jailbreaks, and AgentDojo~\citep{debenedetti2024agentdojo}, InjecAgent~\citep{zhan2024injecagent}, BIPIA~\citep{yi2025bipia}, Agent Security Bench~\citep{zhang2025asb}, and AgentHarm~\citep{andriushchenko2025agentharm} target injection and harmful behavior in applications and agents. Existing benchmarks still have limitations. They measure G3 almost exclusively, and there are no public benchmarks for inference side channels, service integrity, or cost attacks. Attack success is usually decided by an LLM judge, so the results depend on judge prompts and models that change over time. Static attack sets also overestimate robustness. \citet{zhan2025adaptive} used adaptive attacks to bypass all eight indirect injection defenses they tested, with attack success rates above 50\% against each. \citet{nasr2025attacker} used gradient descent, reinforcement learning, random search, and human red teaming to bypass all twelve jailbreak and injection defenses they tested, with attack success rates above 90\% against most of them, whereas the original evaluations of most of these defenses had reported attack success rates near zero. Defense evaluations in MaaS should therefore fix the threat model, give the attacker full knowledge of the defense, and report the compute and query budgets of adaptive attacks, as is done in adversarial example research.

\section{Literature Coding and Findings}\label{sec:findings}

\begin{figure*}[t]
\centering
\includegraphics{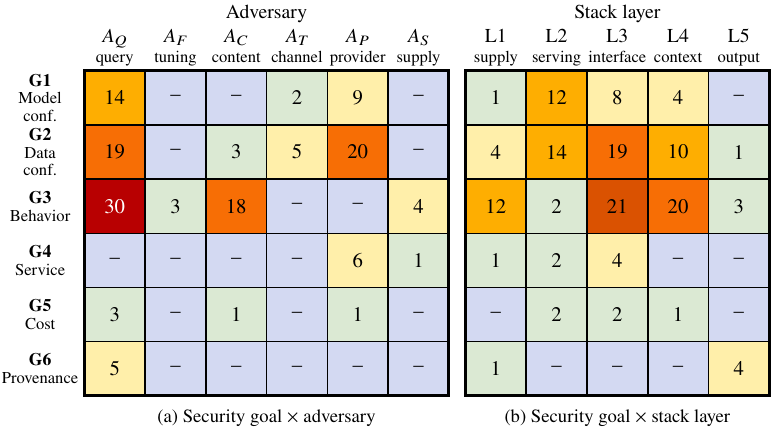}
\caption{The 120 coded studies per security goal by adversary (a) and by stack layer (b)}
\label{fig_4_coding}
\end{figure*}

Figure~\ref{fig_4_coding} shows the distribution of the 120 coded studies over goals, adversaries, and layers. A study is counted in every cell to which it is coded, so the row sums do not equal the corpus size. The field is young: 108 studies (90.0\%) were published in 2023 or later, and 89 (74.2\%) in 2024 or later. The number of studies from 2025 (28) is only half that from 2024 (55), which does not indicate a decline in research: snowballing depends on citations, newer studies have fewer citations, and newer conference papers have not yet all been indexed by the search databases. We report eight findings.

\finding{F1: Research concentrates on behavioral integrity against query-only users.}G3 is the most studied goal (53 studies, 44.2\%), followed by G2 (46 studies, 38.3\%) and G1 (25 studies, 20.8\%). The query-only adversary \adv{Q} appears in 65 studies (54.2\%), and the largest cell in Figure~\ref{fig_4_coding}(a) is \adv{Q} against G3, with 30 studies.

\finding{F2: The goals that define a service are marginal.}Service integrity, availability and cost, and accountability together appear in only 16 studies (13.3\%): 7 for G4, 5 for G5, and 5 for G6 (one of which addresses both G4 and G5). MaaS bills by token, and resource exhaustion attacks shift costs to the paying party, yet none of the 50 defenses targets G5.

\finding{F3: The provider is often the adversary that defenses target but rarely the subject of attack studies.}\adv{P} appears in 20 of the 50 defenses (40.0\%), mainly cryptographic inference, prompt sanitization, and verifiable serving, but in only 6 of the 70 attack and measurement studies (8.6\%). Three of these 6 invert embeddings held by a storage provider, and only 2 (published in 2025 and 2026, respectively) study providers that misreport what they serve~\citep{cai2025substitution,sun2025invisible}. In addition, model equality testing~\citep{gao2025model}, which is coded as a defense by its main contribution, measured deviations from the advertised models on production endpoints. The threat model of a ``dishonest service'' remains almost unstudied.

\finding{F4: Strong guarantees against the provider cannot be deployed by customers.}None of the ten cryptographic or hardware mechanisms against the provider (the eight cryptographic inference protocols, confidential-computing GPUs, and zkLLM, see Table~\ref{tab:defenses}) can be adopted without provider cooperation, and only two of the eight cryptographic inference protocols have been demonstrated on models with 7 billion or more parameters. Of the 20 defenses against \adv{P}, only 6 can be deployed by customers: five prompt sanitization schemes that provide word-level DP or empirical guarantees, and model equality testing.

\finding{F5: Attack and measurement studies are evaluated on real services more than twice as often as defenses.}Of the 70 attack and measurement studies, 45 (64.3\%) include a production service, compared with only 15 of the 50 defenses (30.0\%). Most defenses are evaluated only on open-weight models under the defenders' own control, and whether they transfer to the APIs that most deployments actually use has not been tested.

\finding{F6: Adaptive evaluation is the exception.}Of the 20 defenses targeting G3, 5 (25.0\%) report a full adaptive evaluation, 7 report a partial one, and 8 report none. The adaptive studies of \citet{zhan2025adaptive} and \citet{nasr2025attacker} show the practical consequences of this gap (Section~\ref{sec:evaluation}).

\finding{F7: The inference runtime is a new and weakly protected attack surface.}All four attack and measurement studies involving the side-channel adversary \adv{T} were published in 2024 or 2025, and three of them demonstrated leakage on production systems. SafeKV is the only defense against \adv{T}, and it was evaluated only on open-source serving frameworks, relies on sensitive content detection, and has no adaptive evaluation~\citep{chu2025safekv}. G2 defenses at L2 are dominated by cryptographic inference, which protects against the provider but not against co-located tenants.

\finding{F8: Research on customization, the supply chain, and cross-layer interactions is thin.}\adv{F} appears in 3 studies and \adv{S} in 5 studies, and there is no defense against \adv{S}. Only 7 studies (5.8\%) span more than one layer, and all of them address a single goal. No study in this corpus evaluates how mechanisms at different layers compose, for example whether a TEE-protected runtime with prefix sharing enabled still satisfies G2 against \adv{T}.

\section{Standards and Regulation}\label{sec:standards}

Table~\ref{tab:standards} maps seven groups of frameworks and regulations onto our goals. A framework is rated as covering a goal if it lists concrete threats or requirements for that goal, as partly covering it if it addresses the goal only through general governance or an adjacent category, and as not covering it otherwise. The clauses underlying each rating are as follows. The Open Worldwide Application Security Project (OWASP) Top 10 for LLM Applications~\citep{owasp2025llm} covers G1 and G2 with LLM02 (Sensitive Information Disclosure) and LLM07 (System Prompt Leakage), G3 with LLM01 (Prompt Injection), LLM04 (Data and Model Poisoning), and LLM06 (Excessive Agency), and G5 with LLM10 (Unbounded Consumption), which encompasses denial of wallet and model replication through the API. LLM03 (Supply Chain) addresses only malicious artifacts within G4, so G4 is rated as partly covered. The corresponding techniques in MITRE's Adversarial Threat Landscape for Artificial-Intelligence Systems (ATLAS)~\citep{mitreatlas} include AML.T0024 (Exfiltration via ML Inference API), AML.T0056 (LLM Meta Prompt Extraction), AML.T0057 (LLM Data Leakage), AML.T0051 (LLM Prompt Injection), AML.T0054 (LLM Jailbreak), AML.T0029 (Denial of ML Service), and AML.T0034 (Cost Harvesting), and G4 is likewise only partly covered, by AML.T0010 (ML Supply Chain Compromise). The threat taxonomy and red teaming guide of the Cloud Security Alliance (CSA)~\citep{csa2024llmthreat,csa2025agenticredteam} are engineering-oriented and add agent-specific threats. In its threat taxonomy, insecure supply chain addresses only malicious artifacts within G4 and lack of governance and compliance addresses G6 only indirectly, so both are rated as partly covered. The AI Risk Management Framework of the National Institute of Standards and Technology (NIST) and its Generative AI Profile~\citep{nist2023airmf,nist2024genaiprofile}, ISO/IEC 42001~\citep{iso42001}, and the EU AI Act~\citep{euaiact2024} operate at the level of risk management and obligations. The data privacy and information security risks of the Generative AI Profile correspond to G2 and G3, value chain and component integration addresses only third-party components within G4, model confidentiality is addressed only indirectly under information security risk (G1 partly covered), content provenance has dedicated actions (G6 covered), and resource abuse is not listed as a risk (G5 not covered). Article 15 of the AI Act requires high-risk systems to resist data poisoning, model poisoning, adversarial examples, and confidentiality attacks, Article 55 requires providers of general-purpose models with systemic risk to conduct adversarial testing and ensure cybersecurity protection, and Article 50 requires synthetic content to be marked in a machine-readable format (G6). Articles 15 and 55 apply only to high-risk systems and to general-purpose models with systemic risk, so G1 to G3 are rated as partly covered. Model poisoning in Article 15 targets pretrained components used in training and addresses only malicious artifacts within G4, so G4 is also rated as partly covered. Apart from G6, the controls in Annex A of ISO/IEC 42001 address the goals only at the level of the management system. China's Interim Measures for the Management of Generative Artificial Intelligence Services~\citep{cac2023interim} focus on content security (Article 4), personal information (Article 11), and the labeling of generated content (Article 12), and the national standard GB/T 45654-2025 (Cybersecurity Technology: Basic Security Requirements for Generative Artificial Intelligence Service)~\citep{gbt456542025} additionally specifies requirements for training data security and model security.

The common gap is G4 from the customer's perspective. Supply-chain categories cover malicious artifacts, but no framework requires that customers be able to verify which model processed a request or how many tokens were billed, even though Section~\ref{sec:attacks} has shown that endpoints in practice do deviate from their advertised models~\citep{gao2025model}. G5 shows the opposite situation: OWASP, ATLAS, and CSA all list it as a resource abuse risk, yet none of the coded defenses addresses it.

\begin{table}[t]
\caption{Coverage of the security goals by standards and regulation}
\label{tab:standards}
\footnotesize
\setlength{\tabcolsep}{3pt}
\begin{tabularx}{\linewidth}{@{}Ycccccc@{}}
\toprule
Framework & G1 & G2 & G3 & G4 & G5 & G6 \\
\midrule
OWASP Top 10 for LLM Applications 2025 \citep{owasp2025llm} & \full & \full & \full & \half & \full & \none \\
MITRE ATLAS \citep{mitreatlas} & \full & \full & \full & \half & \full & \none \\
CSA LLM Threats Taxonomy and Agentic AI Red Teaming Guide \citep{csa2024llmthreat,csa2025agenticredteam} & \full & \full & \full & \half & \full & \half \\
NIST AI Risk Management Framework and Generative AI Profile \citep{nist2023airmf,nist2024genaiprofile} & \half & \full & \full & \half & \none & \full \\
ISO/IEC 42001 \citep{iso42001} & \half & \half & \half & \half & \half & \full \\
EU AI Act \citep{euaiact2024} & \half & \half & \half & \half & \none & \full \\
China Interim Measures and GB/T 45654-2025 \citep{cac2023interim,gbt456542025} & \none & \full & \full & \none & \none & \full \\
\bottomrule
\end{tabularx}
\tabnote{\full\ covered, \half\ partly covered, \none\ not covered.}
\end{table}

\section{Lessons and Open Problems}\label{sec:lessons}

\begin{table*}[t]
\caption{Comparison with related surveys and SoKs}
\label{tab:related_surveys}
\footnotesize
\setlength{\tabcolsep}{3pt}
\begin{tabular}{@{}L{0.2\textwidth}C{0.115\textwidth}C{0.115\textwidth}C{0.115\textwidth}C{0.115\textwidth}C{0.115\textwidth}C{0.115\textwidth}@{}}
\toprule
Work & Service stack & Provider and co-tenant adversaries & Service integrity and cost & Deploy\-ability criteria & Literature coding & New experiments \\
\midrule
\citet{papernot2018sok} & \half & \half & \none & \none & \none & \none \\
\citet{liu2023trustworthy} & \none & \none & \none & \none & \none & \none \\
\citet{yao2024llmsecurity} & \none & \none & \none & \none & \half & \none \\
\citet{huang2024trustllm} & \none & \none & \none & \none & \none & \full \\
\citet{das2025security} & \none & \none & \none & \none & \none & \none \\
\citet{ng2023sok} & \none & \half & \none & \half & \half & \none \\
\citet{wang2025supplychain} & \half & \none & \half & \none & \none & \none \\
\citet{deng2025threat} & \half & \none & \none & \none & \none & \none \\
\citet{wang2026sokguardrails} & \none & \none & \none & \half & \none & \full \\
This paper & \full & \full & \full & \full & \full & \none \\
\bottomrule
\end{tabular}
\tabnote{\full\ covered, \half\ partly covered, \none\ not covered.}
\end{table*}

\subsection{Lessons}

\textbf{L1: The provider belongs in the threat model.} Model-centric research treats the provider either as the defender (jailbreaks, stealing) or as a curious party to be excluded by cryptography. Intermediated deployments and opaque billing make the dishonest provider a real threat to G4 (F3), and against this threat customers have only statistical audits such as model equality testing (F4).

\textbf{L2: Guarantees proven at one layer do not survive composition.} The DP of a model no longer holds when deduplication or output filtering leaks membership~\citep{debenedetti2024privacy}. The confidentiality of a model no longer holds when a shared prefix cache leaks timing~\citep{gu2025auditing}. Assurance claims in MaaS must state which layers they cover and which boundaries they assume (F8).

\textbf{L3: In-band defenses have a shelf life.} Defenses that rely on the model (or another model) to distinguish instructions from data, or harmful from benign requests, have mostly been bypassed in third-party adaptive evaluations~\citep{zhan2025adaptive,nasr2025attacker}. Defenses that enforce TB3 structurally~\citep{debenedetti2026camel,xiang2024certifiably} trade flexibility for guarantees that do not depend on model behavior, and this trade-off must be weighed explicitly at design time.

\textbf{L4: Deployability determines impact.} The defenses that customers can deploy (sanitization, guard models, architectural isolation, statistical auditing) all operate on top of an unmodified API. Mechanisms at L1 and L2 depend on the provider's incentives. Research that aims to protect customers should either target L3 to L5 or design mechanisms that let providers prove their own behavior, such as remote attestation and verifiable serving.

\textbf{L5: Measurement coverage is uneven.} Cost abuse, service substitution, and serving-layer leakage are the least studied threats and lack public benchmarks (F2, F7), while existing benchmarks concentrate on jailbreaks (Section~\ref{sec:evaluation}).

\subsection{Open problems}

\textbf{P1: Verifiable serving at production scale.} Eq.~\ref{eq:service} calls for mechanisms that can verify both the identity of a closed-source model and the billed usage at low overhead. Candidate approaches combine TEE attestation of serving images~\citep{apple2024pcc,cai2025substitution}, activation commitments~\citep{ong2025toploc}, and zero-knowledge proofs~\citep{sun2024zkllm}. CoIn begins to cover the metering of hidden tokens~\citep{sun2025coin}, but it requires provider cooperation and cannot confirm that these tokens were generated by the contracted model.

\textbf{P2: Isolation inside the inference runtime.} Prefix sharing, batching, and speculative decoding create cross-tenant channels. SafeKV uses sensitivity detection to decide which cache entries may be shared~\citep{chu2025safekv}, so its protection depends on the detector, which can itself be targeted. This direction needs a definition of tenant isolation for LLM serving that does not rely on content judgments (analogous to constant-time programming), as well as serving engines that trade a measurable loss in cache efficiency for isolation.

\textbf{P3: Practical confidentiality against the provider.} Cryptographic inference remains orders of magnitude slower and requires provider cooperation, and client-side sanitization protects only explicit identifiers. Possible directions include hybrid schemes that use TEEs for nonlinear layers and cryptography for proofs, as well as schemes that split inference between local and remote models with quantifiable leakage bounds.

\textbf{P4: Composable assurance for agents.} Structural defenses enforce TB3 for a single agent. Multi-agent systems, memory, and tool ecosystems need information-flow models in which capabilities, provenance, and permissions compose across components~\citep{debenedetti2026camel,shi2025progent}.

\textbf{P5: Adaptive, service-level evaluation.} Benchmarks should include side-channel, service integrity, and cost attacks, report adaptive attacks with explicit budgets, and run on both production APIs and open models (F5, F6).

\textbf{P6: Defenses against customization and supply-chain attacks.} Fine-tuning APIs need defenses that withstand covert data encoding~\citep{halawi2024covert}, and providers need provenance tracking and scanning for checkpoints and adapters~\citep{zhao2024models}.

\textbf{P7: Accountability that serves both customers and regulators.} Existing provenance mechanisms serve the provider. Auditable logs that record which model, configuration, and policy produced a given output, together with privacy-preserving disclosure of these logs to customers and regulators, could link G4 and G6 and make the obligations in Section~\ref{sec:standards} actionable.

\section{Related Surveys}\label{sec:related}

Table~\ref{tab:related_surveys} compares this paper with related surveys and SoKs. \citet{papernot2018sok} systematized machine learning security and privacy in the MLaaS era and discussed the attack surfaces of the training and inference phases, but this work predates LLMs and their inference runtimes. \citet{liu2023trustworthy} and \citet{huang2024trustllm} organize trustworthiness research by model property and provide evaluation guidelines and benchmarks. \citet{yao2024llmsecurity} and \citet{das2025security} survey LLM security and privacy by type of attack and defense. \citet{ng2023sok} systematized cryptographic neural network computation and treated the server as the adversary, but considered only a curious server and did not address the other layers of the service. \citet{wang2025supplychain} proposed a research agenda for the LLM supply chain, covering issues on the TB5 side of our model. \citet{deng2025threat} surveyed the threats facing AI agents along the agent loop, and \citet{wang2026sokguardrails} systematized jailbreak guardrails and evaluated them with unified experiments. This paper takes the service as the object of modeling, treats the provider and co-located tenants as adversaries, introduces service integrity and cost as goals, evaluates defenses from the perspective of client deployability, and analyzes the literature quantitatively through coding.

\section{Conclusions}\label{sec:conclusion}

MaaS turns the model into a shared, metered service whose security depends on boundaries that model-centric research does not see: the boundary between tenants in the inference runtime, the boundary between inputs of different privilege in the assembled context, and the boundary between customer and provider. We proposed a service-level model with five layers, five trust boundaries, six adversary classes, and six goals, used it to systematize attacks and evaluate 50 assurance mechanisms, and coded 120 studies to locate the field's research effort. The results show that the field has progressed rapidly on jailbreaks and prompt injection, while dishonest services, serving-layer leakage, cost abuse, and the composition of cross-layer protections have received little study, and none of the mechanisms with the strongest guarantees can be deployed unilaterally by customers. The model, evaluation criteria, and appendix coding in this paper can be used to locate these gaps and to state which assurances a MaaS deployment actually provides.

\bibliographystyle{IEEEtranN}
\bibliography{maas}

\appendices

\section{Literature coding}\label{app:coding}

Table~\ref{tab:coding_full} lists the coding of all 120 primary studies with the values defined in Table~\ref{tab:coding}. The last four columns apply to defenses only.

\begin{table*}[p]
\caption{Coding of the 120 primary studies}
\label{tab:coding_full}
\scriptsize
\setlength{\tabcolsep}{1.4pt}
\begin{minipage}[t]{0.49\textwidth}
\begin{tabularx}{\linewidth}[t]{@{}Yllllllllll@{}}
\toprule
Study & Year & T & Layer & Adv. & Goal & Eval. & Guar. & Ad. & Cl. & 7B \\
\midrule
Imitation attacks \citep{wallace2020imitation} & 2020 & A & API & Q & G1 & P & & & & \\
Thieves on Sesame Street \citep{krishna2020thieves} & 2020 & A & API & Q & G1 & O & & & & \\
Grey-box extraction \citep{zanella2021grey} & 2021 & A & API & Q & G1 & O & & & & \\
PromptInject \citep{perez2022ignore} & 2022 & A & Ctx & Q & G1 G3 & P & & & & \\
HouYi \citep{liu2023houyi} & 2023 & A & Ctx & Q & G1 G3 & P & & & & \\
Language model inversion \citep{morris2024inversion} & 2024 & A & API & Q & G1 & O & & & & \\
Logit leakage \citep{finlayson2024logits} & 2024 & A & API & Q & G1 & B & & & & \\
Model stealing \citep{carlini2024stealing} & 2024 & A & API & Q & G1 & B & & & & \\
PLeak \citep{hui2024pleak} & 2024 & A & Ctx & Q & G1 & B & & & & \\
Prompt extraction \citep{zhang2024effective} & 2024 & A & Ctx & Q & G1 & B & & & & \\
KV-cache timing \citep{song2025earlybird} & 2025 & A & Srv & T & G1 G2 & B & & & & \\
LLMmap \citep{pasquini2025llmmap} & 2025 & A & API & Q & G1 & B & & & & \\
Embedding leakage \citep{song2020information} & 2020 & A & Ctx & P & G2 & O & & & & \\
Training data extraction \citep{carlini2021extracting} & 2021 & A & API & Q & G2 & O & & & & \\
GEIA \citep{li2023sentence} & 2023 & A & Ctx & P & G2 & O & & & & \\
Indirect injection \citep{greshake2023not} & 2023 & A & Ctx & C & G2 G3 & B & & & & \\
Neighbourhood membership inference \citep{mattern2023membership} & 2023 & A & API & Q & G2 & O & & & & \\
Personal data leakage \citep{lukas2023analyzing} & 2023 & A & API & Q & G2 & O & & & & \\
Vec2Text \citep{morris2023text} & 2023 & A & Ctx & P & G2 & B & & & & \\
Attribute inference \citep{staab2024beyond} & 2024 & A & API & Q & G2 & B & & & & \\
Min-K\% Prob \citep{shi2024detecting} & 2024 & A & API & Q & G2 & O & & & & \\
Privacy side channels \citep{debenedetti2024privacy} & 2024 & A & Sup Out & Q & G2 & O & & & & \\
RAG privacy \citep{zeng2024good} & 2024 & A & Ctx & Q & G2 & O & & & & \\
SPV-MIA \citep{fu2024practical} & 2024 & A & API & Q & G2 & O & & & & \\
Token-length channel \citep{weiss2024keylogging} & 2024 & A & API & T & G2 & P & & & & \\
Divergence attack \citep{nasr2023scalable} & 2025 & A & API & Q & G2 & B & & & & \\
PromptPeek \citep{wu2025promptpeek} & 2025 & A & Srv & T & G2 & O & & & & \\
Language model red teaming \citep{perez2022red} & 2022 & A & API & Q & G3 & O & & & & \\
GCG \citep{zou2023universal} & 2023 & A & API & Q & G3 & B & & & & \\
Instruction poisoning \citep{wan2023poisoning} & 2023 & A & Sup & S & G3 & O & & & & \\
Jailbroken \citep{wei2023jailbroken} & 2023 & A & API & Q & G3 & P & & & & \\
AutoDAN \citep{liu2024autodan} & 2024 & A & API & Q & G3 & B & & & & \\
Coercing LLMs \citep{geiping2024coercing} & 2024 & A & API & Q & G3 G5 & O & & & & \\
Covert fine-tuning \citep{halawi2024covert} & 2024 & A & Sup & F & G3 & P & & & & \\
Curiosity red teaming \citep{hong2024curiosity} & 2024 & A & API & Q & G3 & O & & & & \\
Harmful fine-tuning \citep{qi2024finetuning} & 2024 & A & Sup & F & G3 & B & & & & \\
Human-feedback backdoor \citep{rando2024universal} & 2024 & A & Sup & S & G3 & O & & & & \\
Injection formalization \citep{liu2024formalizing} & 2024 & A & Ctx & C & G3 & B & & & & \\
Many-shot \citep{anil2024manyshot} & 2024 & A & API & Q & G3 & B & & & & \\
Sleeper agents \citep{hubinger2024sleeper} & 2024 & A & Sup & S & G3 & O & & & & \\
TAP \citep{mehrotra2024tap} & 2024 & A & API & Q & G3 & B & & & & \\
Web-scale poisoning \citep{carlini2024poisoning} & 2024 & A & Sup & S & G3 & P & & & & \\
Adaptive indirect injection attacks \citep{zhan2025adaptive} & 2025 & A & Ctx & C & G3 & O & & & & \\
Adaptive jailbreaks \citep{andriushchenko2025jailbreaking} & 2025 & A & API & Q & G3 & B & & & & \\
Crescendo \citep{russinovich2025crescendo} & 2025 & A & API & Q & G3 & B & & & & \\
PAIR \citep{chao2025jailbreaking} & 2025 & A & API & Q & G3 & B & & & & \\
PoisonedRAG \citep{zou2025poisonedrag} & 2025 & A & Ctx & C & G3 & B & & & & \\
Attacker moves second \citep{nasr2025attacker} & 2026 & A & API Ctx & Q C & G3 & B & & & & \\
Sponge examples \citep{shumailov2021sponge} & 2021 & A & Srv & Q & G5 & B & & & & \\
Engorgio \citep{dong2025engorgio} & 2025 & A & Srv & Q & G5 & O & & & & \\
OverThink \citep{kumar2025overthink} & 2025 & A & Ctx & C & G5 & B & & & & \\
Watermark stealing \citep{jovanovic2024watermark} & 2024 & A & Out & Q & G6 & O & & & & \\
Imitation limits \citep{gudibande2024false} & 2024 & M & API & Q & G1 & B & & & & \\
Prompt-cache audit \citep{gu2025auditing} & 2025 & M & Srv & T & G1 G2 & P & & & & \\
DecodingTrust \citep{wang2023decodingtrust} & 2023 & M & API & Q & G2 G3 & P & & & & \\
Memorization scaling \citep{carlini2023quantifying} & 2023 & M & API & Q & G2 & O & & & & \\
ConfAIde \citep{mireshghallah2024secret} & 2024 & M & Ctx & Q & G2 & B & & & & \\
Disclosures in the wild \citep{mireshghallah2024trust} & 2024 & M & API & P & G2 & P & & & & \\
InjecAgent \citep{zhan2024injecagent} & 2024 & M & Ctx & C & G2 G3 & B & & & & \\
Membership inference reality check \citep{duan2024membership} & 2024 & M & API & Q & G2 & O & & & & \\
\bottomrule
\end{tabularx}
\end{minipage}\hfill
\begin{minipage}[t]{0.49\textwidth}
\begin{tabularx}{\linewidth}[t]{@{}Yllllllllll@{}}
\toprule
Study & Year & T & Layer & Adv. & Goal & Eval. & Guar. & Ad. & Cl. & 7B \\
\midrule
TrustLLM \citep{huang2024trustllm} & 2024 & M & API & Q & G2 G3 & B & & & & \\
AgentDojo \citep{debenedetti2024agentdojo} & 2024 & M & Ctx & C & G3 & B & & & & \\
HarmBench \citep{mazeika2024harmbench} & 2024 & M & API & Q & G3 & B & & & & \\
In-the-wild jailbreaks \citep{shen2024donow} & 2024 & M & API & Q & G3 & B & & & & \\
JailbreakBench \citep{chao2024jailbreakbench} & 2024 & M & API & Q & G3 & B & & & & \\
Malla \citep{lin2024malla} & 2024 & M & API & Q & G3 & P & & & & \\
BIPIA \citep{yi2025bipia} & 2025 & M & Ctx & C & G3 & B & & & & \\
MalHug \citep{zhao2024models} & 2024 & M & Sup & S & G4 & P & & & & \\
Model substitution \citep{cai2025substitution} & 2025 & M & API & P & G4 & O & & & & \\
Hidden tokens \citep{sun2025invisible} & 2026 & M & API & P & G4 G5 & P & & & & \\
Iron \citep{hao2022iron} & 2022 & D & Srv & P & G1 G2 & O & Crypto & n/a & N & N \\
MPCFormer \citep{li2023mpcformer} & 2023 & D & Srv & P & G1 G2 & O & Crypto & n/a & N & N \\
BOLT \citep{pang2024bolt} & 2024 & D & Srv & P & G1 G2 & O & Crypto & n/a & N & N \\
H100 confidential computing \citep{zhu2024hopper} & 2024 & D & Srv & P & G1 G2 & O & HW & n/a & N & Y \\
Instructional fingerprint \citep{xu2024instructional} & 2024 & D & Sup & Q & G1 G6 & O & Stat & P & N & Y \\
Nimbus \citep{li2024nimbus} & 2024 & D & Srv & P & G1 G2 & O & Crypto & n/a & N & N \\
SecFormer \citep{luo2024secformer} & 2024 & D & Srv & P & G1 G2 & O & Crypto & n/a & N & N \\
TransLinkGuard \citep{li2024translinkguard} & 2024 & D & Srv & Q & G1 & O & HW & n/a & N & N \\
BumbleBee \citep{jie2025bumblebee} & 2025 & D & Srv & P & G1 G2 & O & Crypto & n/a & N & Y \\
PUMA \citep{dong2025puma} & 2025 & D & Srv & P & G1 G2 & O & Crypto & n/a & N & Y \\
SHAFT \citep{kei2025shaft} & 2025 & D & Srv & P & G1 G2 & O & Crypto & n/a & N & N \\
SANTEXT+ \citep{yue2021santext} & 2021 & D & API & P & G2 & O & DP & n/a & Y & N \\
DP fine-tuning \citep{yu2022differentially} & 2022 & D & Sup & Q & G2 & O & DP & n/a & N & N \\
DP large-batch \citep{li2022large} & 2022 & D & Sup & Q & G2 & O & DP & n/a & N & N \\
CusText \citep{chen2023customized} & 2023 & D & API & P & G2 & O & DP & n/a & Y & N \\
DP-Forward \citep{du2023dpforward} & 2023 & D & Sup API & Q P & G2 & O & DP & n/a & N & N \\
Hide-and-Seek \citep{chen2023hideandseek} & 2023 & D & API & P & G2 & P & Emp & N & Y & Y \\
PromptPATE \citep{duan2023promptpate} & 2023 & D & Ctx & Q & G2 & P & DP & n/a & Y & Y \\
CoGenesis \citep{zhang2024cogenesis} & 2024 & D & API & P & G2 & B & Emp & N & Y & Y \\
DP-ICL \citep{tang2024privacy} & 2024 & D & Ctx & Q & G2 & B & DP & n/a & Y & Y \\
InferDPT \citep{tong2025inferdpt} & 2025 & D & API & P & G2 & B & DP & n/a & Y & Y \\
SafeKV \citep{chu2025safekv} & 2025 & D & Srv & T & G2 & O & Emp & N & N & Y \\
CaMeL \citep{debenedetti2026camel} & 2026 & D & Ctx & C & G2 G3 & P & Struct & P & Y & Y \\
KV-Cloak \citep{luo2026kvcloak} & 2026 & D & Srv & P & G2 & O & Emp & N & N & Y \\
Baseline defenses \citep{jain2023baseline} & 2023 & D & API & Q & G3 & O & Emp & Y & Y & Y \\
Llama Guard \citep{inan2024llamaguard} & 2023 & D & Out & Q & G3 & O & Emp & N & Y & Y \\
Circuit Breakers \citep{zou2024circuit} & 2024 & D & Sup & Q & G3 & O & Emp & Y & N & Y \\
Erase-and-Check \citep{kumar2025certifying} & 2024 & D & API & Q & G3 & O & Struct & P & Y & Y \\
Instruction hierarchy \citep{wallace2024instruction} & 2024 & D & Sup Ctx & C Q & G3 & P & Emp & N & N & Y \\
Jatmo \citep{piet2024jatmo} & 2024 & D & Sup Ctx & C & G3 & B & Emp & P & Y & Y \\
RAIN \citep{li2024rain} & 2024 & D & Srv & Q & G3 & O & Emp & N & N & Y \\
SafeDecoding \citep{xu2024safedecoding} & 2024 & D & Srv & Q & G3 & O & Emp & N & N & Y \\
Spotlighting \citep{hines2024spotlighting} & 2024 & D & Ctx & C & G3 & P & Emp & N & Y & Y \\
Vaccine \citep{huang2024vaccine} & 2024 & D & Sup & F & G3 & O & Emp & N & N & Y \\
WildGuard \citep{han2024wildguard} & 2024 & D & Out & Q & G3 & O & Emp & N & Y & Y \\
Constitutional Classifiers \citep{sharma2025constitutional} & 2025 & D & Out & Q & G3 & P & Emp & Y & N & Y \\
DataSentinel \citep{liu2025datasentinel} & 2025 & D & Ctx & C & G3 & O & Emp & Y & Y & Y \\
IsolateGPT \citep{wu2025isolategpt} & 2025 & D & Ctx & C & G3 & P & Struct & N & Y & Y \\
Progent \citep{shi2025progent} & 2025 & D & Ctx & C & G3 & B & Struct & P & Y & Y \\
SecAlign \citep{chen2025secalign} & 2025 & D & Sup Ctx & C & G3 & O & Emp & Y & N & Y \\
SmoothLLM \citep{robey2024smoothllm} & 2025 & D & API & Q & G3 & B & Emp & P & Y & Y \\
StruQ \citep{chen2025struq} & 2025 & D & Sup Ctx & C & G3 & O & Emp & P & N & Y \\
RobustRAG \citep{xiang2024certifiably} & 2026 & D & Ctx & C & G3 & O & Struct & P & Y & Y \\
zkLLM \citep{sun2024zkllm} & 2024 & D & Srv & P & G4 & O & Crypto & n/a & N & Y \\
Model equality testing \citep{gao2025model} & 2025 & D & API & P & G4 & P & Stat & n/a & Y & Y \\
TOPLOC \citep{ong2025toploc} & 2025 & D & Srv & P & G4 & O & Stat & n/a & N & Y \\
CoIn \citep{sun2025coin} & 2026 & D & API & P & G4 & O & Stat & n/a & N & Y \\
Green-list watermark \citep{kirchenbauer2023watermark} & 2023 & D & Out & Q & G6 & O & Stat & P & N & N \\
Distortion-free watermark \citep{kuditipudi2024robust} & 2024 & D & Out & Q & G6 & O & Stat & P & N & Y \\
SynthID-Text \citep{dathathri2024scalable} & 2024 & D & Out & Q & G6 & P & Stat & P & N & Y \\
\bottomrule
\end{tabularx}
\end{minipage}
\tabnote{T: type, Eval.: evaluation setting, Guar.: guarantee, Ad.: adaptive evaluation, Cl.: client-deployable, 7B: demonstrated at $\geq$7B. Values as defined in Table~\ref{tab:coding}.}
\end{table*}

\end{document}